\documentclass[runningheads]{llncs}
\usepackage{color}
\usepackage{listings}
\usepackage[misc]{ifsym} 
\usepackage{amsmath}
\usepackage{isabelle,isabellesym}
\usepackage{graphicx}

\def\isadelimproof{}\def\endisadelimproof{}
\def\isatagproof{}\def\endisatagproof{}
\def\isafoldproof{}

\isabellestyle{tt}
\let\endisabellecode=\endisabellecode

\newcommand{\simpl}[1]{{\tt #1}}
\newcommand{\simplarg}[2]{{\tt #1} {\tt #2}}

\begin{document}
\title{A Verified Timsort C Implementation in Isabelle/HOL}
\author{Yu Zhang\inst{1} \and
Yongwang Zhao\inst{1} {\Letter} \and
David Sanan\inst{2}}
\authorrunning{F. Author et al.}
\institute{School of Computer Science and Engineering, Beihang University, Beijing, China
\email{zhaoyw@buaa.edu.cn} \and
School of Computer Science and Engineering, Nanyang Technological University, Singapore}
\maketitle              % typeset the header of the contribution
\begin{abstract}
Formal verification of traditional algorithms are of great significance due to their wide application in state-of-the-art software. 
Timsort is a complicated and hybrid stable sorting algorithm, derived from merge sort and insertion sort. Although Timsort implementation in OpenJDK has been formally verified, there is still not a standard and formally verified Timsort implementation in C programming language. 
This paper studies Timsort implementation and its formal verification using a generic imperative language - Simpl in Isabelle/HOL. Then, we manually generate an C implementation of Timsort from the verified Simpl specification. Due to the C-like concrete syntax of Simpl, the code generation is straightforward. The C implementation has also been tested by a set of random test cases. 

%Formally verified programs are of higher reliability than tested programs. Program verification used to be expensive but with the development of verification tool, like Isabelle/HOL, it has become manageable to carry out. We investigate Timsort using Isabelle/HOL, which is widely used in different programming languages. We first implement Timsort using Simpl, a generic imperative language model. Then formal verification is used to prove the invariants. Finally, we generated executable verified Timsort implementation in C based on Simpl specification.

\end{abstract}
\section{Introduction}\label{sec:intro}

Formal verification has been considered as a promising way to the reliability of programs. With development of verification tools, it is possible to perform fully formal verification of large and complex programs in recent years \cite{Klein09,Leroy09}. Formal verification of traditional algorithms are of great significance due to their wide application in state-of-the-art software. 
The goal of this paper is the functional verification of sorting algorithms as well as generation of C source code. We investigated Timsort algorithm which is a hybrid stable sorting algorithm, derived from merge sort and insertion sort, designed to work well on many kinds of real-world data. 

Tim Peters invented Timsort algorithm and applied it in the Python standard library. Afterwards it has also been used to sort arrays of non-primitive type in Java, on the Android platform and in GNU Octave. 
Gouw et al.\cite{deGouw2015OpenJDKsCase} have carried our formal verification of the OpenJDK's Timsort implementation using KeY. Timsort is the main sorting algorithm provided by the Java standard library. KeY is a Java verification tool and a semi-automatic, interactive theorem prover, covering nearly full sequential Java. However, there is still not a standard and formally verified Timsort implementation in C programming language. Tim Peters \cite{PetersTimsortC} himself released a C version Timsort, which is actually a part of the implementation of Python's List data structure in C and has not been verified.

This paper studies Timsort implementation and its formal verification using Simpl \cite{SchirmerVerificationHOL} in Isabelle/HOL. Different from KeY which mainly focuses on Java programs, Simpl is a generic imperative language embedded into Isabelle/HOL that was designed as an intermediate language for program verification. 
In Isabelle/HOL, the GHC's sorting algorithm for lists has been formalized and its correctness and stability have been proved in \cite{Sternagel2013}. Quicksort algorithm has been verified in Simpl based on split heap model\cite{Simpl-AFP}, which cost only less than 100 interactive proofs. 
Simpl has been deeply used in formal verification of the seL4 OS kernel \cite{Klein09}, where the C source code of the kernel is automatically translated into Simpl specification by the verified tools, CParser and Autocorres. 
Lars Noschinski et al.\cite{RizkallahAVerification} has formally verified a certifying algorithm checkers for connectedness of graphs written in C from the library LEDA using Simpl and Autocorres. Besides, in order to reason about concurrent programs, Sanan et al. extends Simpl to CSimpl \cite{Sanan2017CSimpl:Programs} which is an extension of Simpl with concurrency-oriented language features and verification techniques. 

In this paper, we specify the Timsort algorithm using Simpl in the Isabelle/HOL theorem prover, and then generate real C source code after its functional verification \footnote{The Isabelle/HOL specification and proof, and the generated C code are available at \url{https://github.com/LVPGroup/TimSort/}}. As a first step, the C code generation is done manually according to the Simpl specification. Thanks to the C-like concrete syntax of Simpl in Isabelle/HOL, the generation is straightforward and could be easily implemented by a translator in the future. 
Compare to the post-hoc verification of Timsort algorithms within KeY \cite{deGouw2015OpenJDKsCase}, we use Simpl and Isabelle/HOL to specify and verify Timsort algorithm providing machine-checked proof, and then export the specification into C code. 
Second, KeY is a proof assistant designed for Java programs, whilst Isabelle/HOL and Simpl are more general. Therefore, it is possible to generate verified a verified Timsort implementation in other imperative languages. Third, Simpl is embedded in Isabelle/HOL and we can make use of its comprehensive libraries and stronger solvers/provers. So we expect the process of verification comes at a lower cost compared with its counterpart in KeY.

\section{Preliminaries}
\subsection{Timsort Algorithm}
Timsort  is an effective combination of merge sort and insertion sort, which subtly modifies the two classical algorithms to reach better performance. It is a stable sort with complexity of $O(n\log n)$ at worst case and $O(n)$ at best. Timsort is designed to take advantage of partial ordering that already exists in the data so it's remarkably fast for nearly sorted data sequence and reverse sorted data. The procedure of Timsort basically follows the pattern of divide-and-conquer:

\begin{itemize}
\item Divide an input array into sub-arrays with a minimal length $min\_run$
\item Sort each sub-array by binary sort(a combination of binary search and insertion sort)
\item Merge all the sorted sub-arrays into a single array using a modified merge sort
\end{itemize}

The keys of Timsort lie in the details of these steps. We refer to the sub-array as $run$ and the minimal length of runs as $min\_run$. 
The first step is to calculate the parameter $min\_run$. It should not be too large, because insertion sort is only effective for short arrays. It also should not be too small, because it will lead to more merge iterations in the last step. Based on experiments the values work well between 32 and 65. Besides, the optimal value is when $N/min\_run$ is a power of 2 where $N$ is the length of input array because merge sort works perfectly on balanced sub-arrays. But there is not always such an integer $min\_run$ for every possible value of $N$, so we pick a value in range (32,65) that $N/min\_run$ is a power of 2 or is strictly less than a power of 2. 

The second step is to divide the input array into $run$s. We first count the number of continuous increasing of decreasing elements from current pointer. If the number is greater than $min\_run$, then this sorted sub-array will be count as a $run$ and if it's decreasing reverse it in place. Otherwise, we extend this sub-array to the length of $min\_run$ and using binary sort to keep it sorted. 

The last step is to merge all these sorted sub-arrays. It is always wise to merge the sub-arrays of equal or similar size. To achieve this, Timsort uses stacks to store the indexes and lengths of these sub-arrays. Every time a new $run$ is created, the length of the $run$ will be pushed to the stack $run\_len$ and its index to the stack $run\_base$. More importantly, if consider the top three elements in stack $run\_len$ are X, Y and Z, the stack maintains two invariants: $X > Y + Z$ and $Y>Z$. 

These invariants aim at maintaining run lengths as close to each other as possible to ensure balanced merges, which are more efficient. Once a new element is pushed and the rules are broken, Y will be merged with the smaller one between X and Z. The merging continues until both the invariants are satisfied. After the complete input array has been divided into $run$s, the top two $run$s in the stack are merged until there is only one $run$ remains, which is the sorted array. Consider the lengths of $run$s in the stack are: 128, 64, 32, 16, 8, 4, 2 and finally the last sub-array comes with length of 2. Then there will be 7 perfectly balanced merges. 
% * <zhaoyongwang@gmail.com> 2018-07-02T01:17:02.243Z:
% 
% 这小节大概介绍一下Timsort算法，及它复杂的地方。
% 
% ^ <zhaoyongwang@gmail.com> 2018-07-02T01:37:53.009Z.

\subsection{Simpl in Isabelle/HOL}

Schirmer introduces in~\cite{SchirmerVerificationHOL} a verification framework for imperative sequential programs developed in Isabelle/HOL. 
The verification framework includes a generic imperative language, called Simpl, which is composed of the necessary constructors to capture most of the features present in common sequential languages, such as conditional branching, loops, abrupt termination and exceptions, assertions, mutually recursive functions, expressions with side effects, and nondeterminism. Additionally, Simpl can express memory related features like the memory heap, pointers, and pointers to functions. The Simpl verification framework also includes a Floyd/Hoare-like logic to reason about partial and total correctness, and on top of it, the framework implements a verification condition generator (VCG) to ease the verification process.

%In this section, we introduce the syntax and semantics of Simpl language model which supports most of the imperative programming language features. To illustrate this, we use some language features in C as examples and explain the implementation of these typical features. Moreover, we will introduce Hoare Logic in Simpl, which is the basic theory of program reasoning, and verification condition generator, which is the automatic tool to simplify the Hoare Logic formula to high order logic.\\
%Simpl provides a verification environment for imperative programming languages. It does not aim at any specific languages but most real programming languages like C, Pascal or Java can be embedded in its general language model. Simpl is of high flexibility because it only defines the control flow statements. Its syntax is defined as below:\\
The syntax of Simpl (shown in Fig.~\ref{fig:Syntax}) is defined in terms of states, of type {\tt 's}; a set of fault types, of type {\tt 'f}; and a set of procedure names of type {\tt 'p}. 
The constructor {\tt Skip} indicates program termination; {\simplarg {Seq} {s1 s2}, {\simplarg {Cond} {b c1 c2}}, and {\simplarg {While} {b c}} are respectively the standard constructors for sequential, conditional, and loop statements. \simpl{Throw} and {\simplarg {Throw} {c1 c2}} are the complements  for abrupt termination of programs of \simpl{Skip} and {\simplarg {Seq} {c1 c2}}, and they allow to model exceptions. {\simplarg {Call} p} invokes procedure {\tt p}; {\simplarg {Guard} {f g c}} represents assertions, where {\emph c} is executed if the guard {\tt g} holds in the current state, fault of type {\tt 'f} is raised otherwise. Finally, \simplarg{Spec}{r} introduces a nondeterministic behavior expressed by relation \simpl{r}, and \simplarg{DynCom}{cs} provides state dependent dynamic command transformation using the function \simpl{cs} which is used to model blocks and functions with arguments. 
The function call in Simpl is implemented by the dynamic command.

%$\Gamma$, which is a partial function from the set {\tt 'p} of procedure names to the body of the procedures. 
\begin{figure}[t]
	\begin{isabellebody} 
		\footnotesize
		\isamarkuptrue%
		\isacommand{type{\isacharunderscore}synonym}\isamarkupfalse%
		\ {\isacharprime}s\ bexp\ {\isacharequal}\ {\isachardoublequoteopen}{\isacharprime}s\ set{\isachardoublequoteclose}\isanewline
		\isacommand{datatype}\isamarkupfalse%
		\ {\isacharparenleft}{\isacharprime}s{\isacharcomma}\ {\isacharprime}p{\isacharcomma}\ {\isacharprime}f{\isacharparenright}\ com\ {\isacharequal}\isanewline
		\ \ Skip {\isacharbar}\ Throw {\isacharbar}\ Basic\ {\isachardoublequoteopen}{\isacharprime}s\ {\isasymRightarrow}\ {\isacharprime}s{\isachardoublequoteclose} {\isacharbar}\ Spec\ {\isachardoublequoteopen}{\isacharparenleft}{\isacharprime}s\ {\isasymtimes}\ {\isacharprime}s{\isacharparenright}\ set{\isachardoublequoteclose} {\isacharbar}\ Call\ {\isachardoublequoteopen}{\isacharprime}p{\isachardoublequoteclose}\isanewline
		\ \ {\isacharbar}\ Seq\ {\isachardoublequoteopen}{\isacharparenleft}{\isacharprime}s\ {\isacharcomma}{\isacharprime}p{\isacharcomma}\ {\isacharprime}f{\isacharparenright}\ com{\isachardoublequoteclose}\ {\isachardoublequoteopen}{\isacharparenleft}{\isacharprime}s{\isacharcomma}{\isacharprime}p{\isacharcomma}\ {\isacharprime}f{\isacharparenright}\ com{\isachardoublequoteclose}\ \ \ \ \isanewline
		\ \ {\isacharbar}\ Cond\ {\isachardoublequoteopen}{\isacharprime}s\ bexp{\isachardoublequoteclose}\ {\isachardoublequoteopen}{\isacharparenleft}{\isacharprime}s{\isacharcomma}{\isacharprime}p{\isacharcomma}{\isacharprime}f{\isacharparenright}\ com{\isachardoublequoteclose}\ \ {\isachardoublequoteopen}{\isacharparenleft}{\isacharprime}s{\isacharcomma}{\isacharprime}p{\isacharcomma}{\isacharprime}f{\isacharparenright}\ com{\isachardoublequoteclose}\isanewline
		\ \ {\isacharbar}\ While\ {\isachardoublequoteopen}{\isacharprime}s\ bexp{\isachardoublequoteclose}\ {\isachardoublequoteopen}{\isacharparenleft}{\isacharprime}s{\isacharcomma}{\isacharprime}p{\isacharcomma}{\isacharprime}f{\isacharparenright}\ com{\isachardoublequoteclose} {\isacharbar}\ DynCom\ {\isachardoublequoteopen}{\isacharprime}s\ {\isasymRightarrow}\ {\isacharparenleft}{\isacharprime}s{\isacharcomma}{\isacharprime}p{\isacharcomma}{\isacharprime}f{\isacharparenright}\ com{\isachardoublequoteclose}\ \isanewline
		\ \ {\isacharbar}\ Guard\ {\isachardoublequoteopen}{\isacharprime}f{\isachardoublequoteclose}\ {\isachardoublequoteopen}{\isacharprime}s\ bexp{\isachardoublequoteclose}\ {\isachardoublequoteopen}{\isacharparenleft}{\isacharprime}s{\isacharcomma}{\isacharprime}p{\isacharcomma}{\isacharprime}f{\isacharparenright}\ com{\isachardoublequoteclose}\ \isanewline	
		\ \ {\isacharbar}\ Catch\ {\isachardoublequoteopen}{\isacharparenleft}{\isacharprime}s{\isacharcomma}{\isacharprime}p{\isacharcomma}{\isacharprime}f{\isacharparenright}\ com{\isachardoublequoteclose}\ {\isachardoublequoteopen}{\isacharparenleft}{\isacharprime}s{\isacharcomma}{\isacharprime}p{\isacharcomma}{\isacharprime}f{\isacharparenright}\ com{\isachardoublequoteclose}
%\isanewline
		
%		\isacommand{datatype}\isamarkupfalse%
%		\ {\isacharparenleft}{\isacharprime}s{\isacharcomma}{\isacharprime}f{\isacharparenright}\ xstate\ {\isacharequal}\ Normal\ {\isacharprime}s\ {\isacharbar}\ Abrupt\ {\isacharprime}s\ {\isacharbar}\ Fault\ {\isacharprime}f\ {\isacharbar}\ Stuck
%		
%		\isacommand{type{\isacharunderscore}synonym}\isamarkupfalse%
%		{\isacharparenleft}{\isacharprime}s{\isacharcomma}{\isacharprime}p{\isacharcomma}{\isacharprime}f{\isacharparenright}\ config\ {\isacharequal}\ {\isachardoublequoteopen}{\isacharparenleft}{\isacharprime}s{\isacharcomma}{\isacharprime}p{\isacharcomma}{\isacharprime}f{\isacharparenright}com\ \ {\isasymtimes}\ {\isacharparenleft}{\isacharprime}s{\isacharcomma}{\isacharprime}f{\isacharparenright}\ xstate{\isachardoublequoteclose}
%		
%		\isacommand{type{\isacharunderscore}synonym}\isamarkupfalse%
%		\ {\isacharparenleft}{\isacharprime}s{\isacharcomma}{\isacharprime}p{\isacharcomma}{\isacharprime}f{\isacharparenright}\ body\ {\isacharequal}\ {\isachardoublequoteopen}{\isacharprime}p\ {\isasymRightarrow}\ {\isacharparenleft}{\isacharprime}s{\isacharcomma}{\isacharprime}p{\isacharcomma}{\isacharprime}f{\isacharparenright}\ com\ option{\isachardoublequoteclose}		
%		
	\end{isabellebody}%	
	\caption{Abstract Syntax of Simpl Language}\label{fig:Syntax}
	\vspace{-10px}
\end{figure}

Based on operational semantics, Simpl implement a Hoare proof system for functional correctness of programs. In Simpl, the specification of Hoare logic has the form:
$$\Gamma,\Theta \vdash _{/F}\ P\ c\ Q,A$$
$\Gamma$ is the procedure environment. $\Theta$ is a set of assumptions that contains the specifications we can utilize while verifying the program $c$. $P,Q,A$ are the precondition, postcondition for normal termination and abrupt termination respectively. Both partial and total correctness are defined inductively in Simpl. Moreover, both of them are proved sound and complete with reference to their semantics. 
The main tool in Simpl to utilize Hoare logic investigating programs is a verification condition generator that is implemented as tactic called $vcg$. For a specification $\Gamma,\Theta\vdash P\ c\ Q,A$, applying $vcg$ simplifies the problem to the form $P \subseteq WP$, where $WP$ is the weakest precondition after execution of $c$, $Q$ and $A$. 

Here, we illustrate how to specify and verify programs in Simpl. First, we use the keyword ``procedures'' to define a procedure and specify its signature and body as follows. 

\begin{isabellebody}
\isacommand{procedures}\isamarkupfalse%
\ \ Fac\ {\isacharparenleft}N{\isacharcolon}{\isacharcolon}nat\ {\isacharbar}\ R{\isacharcolon}{\isacharcolon}nat{\isacharparenright}\ \isanewline
\ \ {\isachardoublequoteopen}IF\ {\isasymacute}N\ {\isacharequal}\ {\isadigit{0}}\ THEN\ {\isasymacute}R\ {\isacharcolon}{\isacharequal}{\isacharequal}\ {\isadigit{1}}\isanewline
\ \ \ ELSE\ {\isasymacute}R\ {\isacharcolon}{\isacharequal}{\isacharequal}\ CALL\ Fac{\isacharparenleft}{\isasymacute}N\ {\isacharminus}\ {\isadigit{1}}{\isacharparenright}{\isacharsemicolon}{\isacharsemicolon}\isanewline
\ \ \ \ \ \ \ \ {\isasymacute}R\ {\isacharcolon}{\isacharequal}{\isacharequal}\ {\isasymacute}N\ {\isacharasterisk}\ {\isasymacute}R
\isanewline \ \ \ FI{\isachardoublequoteclose}%
\end{isabellebody}

Then, we use pre- and post-condition to define its correctness specification and use Hoare logic to prove its correctness. We prove the specification to show the correctness of the procedure. First, we apply the rule $HoarePartial.ProcRec1$ to expand the body of procedure. Then, the method $vcg$ reduces the problem to the level of first order logic. Finally, we solve it automatically.

\begin{isabellebody}

\isacommand{lemma}\isamarkupfalse%
\ {\isacharparenleft}\isakeyword{in}\ Fac{\isacharunderscore}impl{\isacharparenright}\ Fac{\isacharunderscore}spec{\isacharcolon}\isanewline
\isakeyword{shows}\ {\isachardoublequoteopen}{\isasymforall}n{\isachardot}\ {\isasymGamma}{\isasymturnstile}\ {\isasymlbrace}{\isasymacute}N\ {\isacharequal}\ n{\isasymrbrace}\ {\isasymacute}R\ {\isacharcolon}{\isacharequal}{\isacharequal}\ PROC\ Fac{\isacharparenleft}{\isasymacute}N{\isacharparenright}\ {\isasymlbrace}{\isasymacute}R\ {\isacharequal}\ fac\ n{\isasymrbrace}{\isachardoublequoteclose}\isanewline
\isadelimproof
\endisadelimproof
\isatagproof
\isacommand{apply}\isamarkupfalse%
\ {\isacharparenleft}hoare{\isacharunderscore}rule\ HoarePartial{\isachardot}ProcRec{\isadigit{1}}{\isacharparenright}\isanewline
\isacommand{apply}\isamarkupfalse%
\ vcg
\ \isacommand{apply}\ clarsimp
\ \isacommand{apply}\ {\isacharparenleft}case{\isacharunderscore}tac\ N{\isacharparenright}
\ \isacommand{apply}\ auto \ \isacommand{done}\isamarkupfalse%
\endisatagproof

\end{isabellebody}
So far we have got the specification of factorial procedure and now we are able to make use of it. The syntax for procedure call is straightforward and when reaching a procedure call, verification condition generator looks for the specification and applies the rule $HoarePartial.ProcSpec$ instantiated with the specification.
\begin{isabellebody}
\isacommand{lemma}\isamarkupfalse%
\ {\isacharparenleft}\isakeyword{in}\ Fac{\isacharunderscore}impl{\isacharparenright}\isanewline
\isakeyword{shows}\ {\isachardoublequoteopen}{\isasymGamma}{\isasymturnstile}\ {\isasymlbrace}{\isasymacute}N\ {\isacharequal}\ {\isadigit{3}}{\isasymrbrace}\ {\isasymacute}R\ {\isacharcolon}{\isacharequal}{\isacharequal}\ CALL\ Fac{\isacharparenleft}{\isasymacute}N{\isacharparenright}\ {\isasymlbrace}\ {\isasymacute}R\ {\isacharequal}\ {\isadigit{6}}\ {\isasymrbrace}{\isachardoublequoteclose}\isanewline
\isadelimproof
\ \ %
\endisadelimproof
\isatagproof
\isacommand{apply}\isamarkupfalse%
\ vcg\ \isacommand{apply} {\isacharparenleft}auto\ simp\ add{\isacharcolon}numeral{\isacharunderscore}{\isadigit{3}}{\isacharunderscore}eq{\isacharunderscore}{\isadigit{3}}{\isacharparenright}\ \isacommand{done}\isamarkupfalse%
\endisatagproof
\end{isabellebody}

% * <zhaoyongwang@gmail.com> 2018-07-02T01:15:36.947Z:
% 
% 这部分需要突出Simpl支持C语言各方面的能力（比如动态过程调用、递归调用、指针、位运算等），Simpl的Hoare Logic等验证，及VCG的自动化等。参考[2006]Verification of Sequential Imperative Programs in Isabelle/HOL_PhD Thesis
% 
% ^ <yu.zhang.7161@student.lu.se> 2018-07-07T10:04:18.515Z.
\section{Timsort Implementation by Simpl}

As mentioned in Section~\ref{sec:intro}, the first step is to specify Timsort algorithm using Simpl. We develop Simpl specification of Timsort according to the OpenJDK's implementation. Thanks to the expressiveness of Simpl, the specification is a direct mapping from the implementation for most of the statements. However, some features of Java are different from the general language model, making necessary to introduce additional Simpl specification to model them. In summary, we specify all functions of Timsort as show in Fig. \ref{fig:timsort_spec}. 

\begin{figure}[t]
\begin{center}
\includegraphics[width=\textwidth]{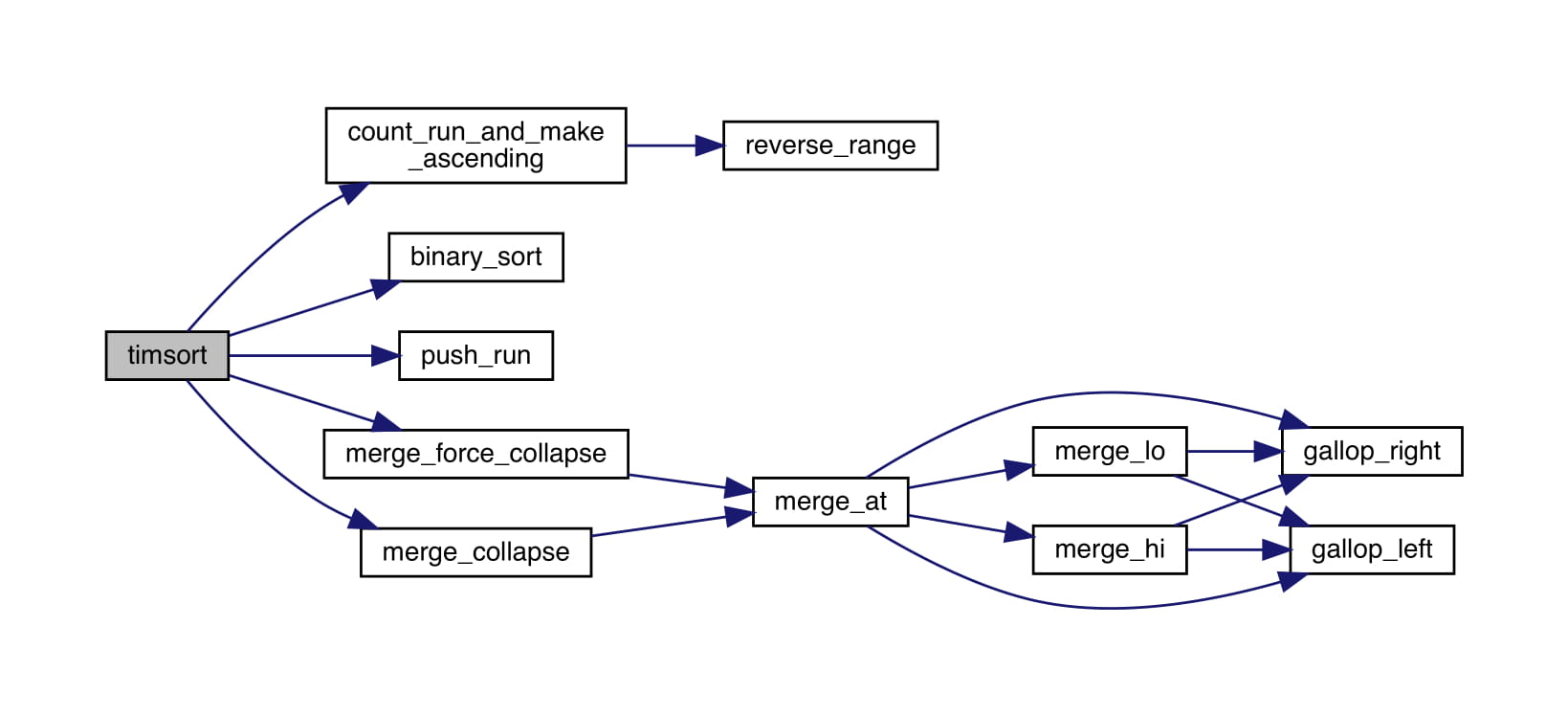}
\end{center}
\caption{Timsort functions and their call graph implemented in Simpl Specification}
\label{fig:timsort_spec}
\end{figure}

\subsection{Instance Variables}

Java is an object-oriented programming language where data and methods are encapsulated into classes. The Timsort algorithm is implemented as a class in Java, so it has its own instance variables and methods. In Simpl, we use global variables to model Timsort instance variables, and procedures to model the class methods. Simpl uses \textbf{hoarestate} to store variables and the hoarestate that contains global variables begins with prefix ``globals-''. The declaration is as below:

\begin{isabellebody}
\isacommand{hoarestate}\isamarkupfalse%
\ globals{\isacharunderscore}var\ {\isacharequal}\ \isanewline
\ stack{\isacharunderscore}size{\isacharcolon}{\isacharcolon}\ nat\isanewline
\ run{\isacharunderscore}base\ {\isacharcolon}{\isacharcolon}\ {\isachardoublequoteopen}nat\ list{\isachardoublequoteclose}\isanewline
\ run{\isacharunderscore}len\ {\isacharcolon}{\isacharcolon}\ {\isachardoublequoteopen}nat\ list{\isachardoublequoteclose}\isanewline
\ stack{\isacharunderscore}len\ {\isacharcolon}{\isacharcolon}\ nat\ \isanewline
\ a\ {\isacharcolon}{\isacharcolon}\ {\isachardoublequoteopen}int\ list{\isachardoublequoteclose}\ \isanewline
\ global{\isacharunderscore}min{\isacharunderscore}gallop\ {\isacharcolon}{\isacharcolon}\ nat\isanewline
\end{isabellebody}

Most of variables are declared in the same way as the Java implementation except the stacks and its size. Because both stacks $run\_base$ and $run\_len$ store non-negative elements and arrays are modeled as lists in Simpl, they have the type ``nat list'', which means list of natural numbers. Similarly, the variable $stack\_size$ is the size of the two stacks so it is defined as type $nat$. Isabelle/HOL itself has the type int, but the advantage of $nat$ over $int$ is that many auxiliary definitions and lemmas are defined using natural numbers and their inductive structures. As a result, we can use these useful definitions and lemmas for free.

\subsection{Restate Methods in Simpl}

\subsubsection{System methods}

The Timsort implementation in Java involves copying a part of an array from source position to destinate position when doing binary sort and merge sort in gallop mode. This is achieved by the system methods System.arraycopy() in Java. This method is a native method, which means it is written in other programming language and may be executed differently on different architectures and virtual machines. In most programming languages, the function to copy memory are provided in standard library, which is assumed to be correct for this stage. So we define this method on the Isabelle/HOL level, which is directly used in the Simpl specification and from the view of Simpl it looks like a method from the ``library''. Over the Isabelle/HOL specification we prove properties of the system method that are necessary for the correctness of the Simpl specification.  Moreover, additional lemmas can be proven easily for future uses. The definition for copying $ys[m..m+l]$ to $xs[n..n+l]$ is as below:

\begin{isabellebody}
\isacommand{definition}\isamarkupfalse%
\ list{\isacharunderscore}copy\ {\isacharcolon}{\isacharcolon}\ {\isachardoublequoteopen}{\isacharprime}a\ list\ {\isasymRightarrow}\ nat\ {\isasymRightarrow}\ {\isacharprime}a\ list\ \ {\isasymRightarrow}\ nat\ {\isasymRightarrow}\ nat\ {\isasymRightarrow}\ {\isacharprime}a\ list{\isachardoublequoteclose}\ \isakeyword{where}\isanewline
{\isachardoublequoteopen}list{\isacharunderscore}copy\ xs\ n\ ys\ m\ l\ {\isacharequal}\ {\isacharparenleft}take\ n\ xs{\isacharparenright}\ {\isacharat}\ {\isacharparenleft}take\ l\ {\isacharparenleft}drop\ m\ ys{\isacharparenright}{\isacharparenright}\ {\isacharat}\ {\isacharparenleft}drop\ {\isacharparenleft}n{\isacharplus}l{\isacharparenright}\ xs{\isacharparenright}{\isachardoublequoteclose}\isanewline
\end{isabellebody}

And because \textit{IndexOutOfBoundException} is a throw in Java when source position plus copy length exceeds the length of source array, or when destinate position plus copy length exceeds the length of destinate array, we can use these constraints as assumptions to conclude the correctness of our definition and some useful lemmas. Here we prove that in the result array elements are preserved with reference to original arrays and the length of array does not change.

\begin{isabellebody}
\isacommand{lemma}\isamarkupfalse%
\ list{\isacharunderscore}copy{\isacharunderscore}len{\isacharbrackleft}simp{\isacharbrackright}{\isacharcolon}{\isachardoublequoteopen}{\isacharparenleft}m{\isacharplus}l{\isacharparenright}{\isasymle}length\ ys\ {\isasymLongrightarrow}\ {\isacharparenleft}n{\isacharplus}l{\isacharparenright}{\isasymle}length\ xs\ {\isasymLongrightarrow}\ \isanewline \ \ \ \ \ \ \ \ \ \ \ \ \ \ \ 
 {\isacharparenleft}length\ {\isacharparenleft}list{\isacharunderscore}copy\ xs\ n\ ys\ m\ l{\isacharparenright}\ {\isacharequal}\ length\ xs{\isacharparenright}{\isachardoublequoteclose}\isanewline
\isadelimproof
\ \ %
\endisadelimproof
\isatagproof
\isacommand{by}\isamarkupfalse%
\ {\isacharparenleft}auto\ simp\ add{\isacharcolon}list{\isacharunderscore}copy{\isacharunderscore}def{\isacharparenright}%
\endisatagproof
\isanewline
\isanewline
\isacommand{lemma}\isamarkupfalse%
\ list{\isacharunderscore}copy{\isacharunderscore}i{\isacharunderscore}front{\isacharbrackleft}simp{\isacharbrackright}{\isacharcolon}{\isachardoublequoteopen}{\isacharparenleft}n{\isacharplus}l{\isacharparenright}{\isasymle}length\ xs\ {\isasymLongrightarrow}\ {\isacharparenleft}m{\isacharplus}l{\isacharparenright}{\isasymle}length\ ys\ {\isasymLongrightarrow}\isanewline \ \ \ \ \ \ \ \ \ 
 i{\isacharless}n\ {\isasymLongrightarrow}\ {\isacharparenleft}list{\isacharunderscore}copy\ xs\ n\ ys\ m\ l{\isacharparenright}{\isacharbang}i\ {\isacharequal}\ xs{\isacharbang}i{\isachardoublequoteclose}\isanewline
\isadelimproof
\ \ %
\endisadelimproof
\isatagproof
\isacommand{by}\isamarkupfalse%
\ {\isacharparenleft}auto\ simp\ add{\isacharcolon}list{\isacharunderscore}copy{\isacharunderscore}def{\isacharparenright}
\endisatagproof
{\isafoldproof}%
\isadelimproof
\isanewline
\endisadelimproof
\isanewline
\isacommand{lemma}\isamarkupfalse%
\ list{\isacharunderscore}copy{\isacharunderscore}i{\isacharunderscore}mid{\isacharbrackleft}simp{\isacharbrackright}{\isacharcolon}{\isachardoublequoteopen}{\isacharparenleft}n{\isacharplus}l{\isacharparenright}{\isasymle}length\ xs\ {\isasymLongrightarrow}\ {\isacharparenleft}m{\isacharplus}l{\isacharparenright}{\isasymle}length\ ys\isanewline \ \ \ \ \ \ \ \ \ 
 {\isasymLongrightarrow}\ i{\isasymge}n{\isasymand}i{\isacharless}{\isacharparenleft}n{\isacharplus}l{\isacharparenright}\ {\isasymLongrightarrow}\ {\isacharparenleft}list{\isacharunderscore}copy\ xs\ n\ ys\ m\ l{\isacharparenright}{\isacharbang}i\ {\isacharequal}\ ys{\isacharbang}{\isacharparenleft}i{\isacharminus}n{\isacharplus}m{\isacharparenright}{\isachardoublequoteclose}\isanewline
\isadelimproof
\ \ %
\endisadelimproof
\isatagproof
\isacommand{apply}\isamarkupfalse%
\ {\isacharparenleft}auto\ simp\ add{\isacharcolon}list{\isacharunderscore}copy{\isacharunderscore}def{\isacharparenright}\isanewline
\ \ \isacommand{apply}\isamarkupfalse%
\ {\isacharparenleft}subgoal{\isacharunderscore}tac\ {\isachardoublequoteopen}min\ {\isacharparenleft}length\ xs{\isacharparenright}\ n\ {\isacharequal}\ n{\isachardoublequoteclose}{\isacharparenright}\isanewline
\ \ \ \isacommand{apply}\isamarkupfalse%
\ {\isacharparenleft}simp{\isacharparenright}
\ \ \ \isacommand{apply}\isamarkupfalse%
\ {\isacharparenleft}subgoal{\isacharunderscore}tac\ {\isachardoublequoteopen}i{\isacharminus}n\ {\isacharless}\ l{\isachardoublequoteclose}{\isacharparenright}\isanewline
\ \ \ \ \isacommand{by}\isamarkupfalse%
\ {\isacharparenleft}auto\ simp\ add{\isacharcolon}add{\isachardot}commute{\isacharparenright}
\endisatagproof
{\isafoldproof}%
\isadelimproof
\isanewline
\endisadelimproof
\ \ \isanewline
\isacommand{lemma}\isamarkupfalse%
\ list{\isacharunderscore}copy{\isacharunderscore}i{\isacharunderscore}end{\isacharbrackleft}simp{\isacharbrackright}{\isacharcolon}{\isachardoublequoteopen}{\isacharparenleft}n{\isacharplus}l{\isacharparenright}{\isasymle}length\ xs\ {\isasymLongrightarrow}\ {\isacharparenleft}m{\isacharplus}l{\isacharparenright}{\isasymle}length\ ys\ {\isasymLongrightarrow}\isanewline \ \ \ \ 
 i{\isasymge}n{\isacharplus}l{\isasymand}i{\isacharless}length\ xs\ {\isasymLongrightarrow}\ {\isacharparenleft}list{\isacharunderscore}copy\ xs\ n\ ys\ m\ l{\isacharparenright}{\isacharbang}i\ {\isacharequal}\ xs{\isacharbang}i{\isachardoublequoteclose}\isanewline
\isadelimproof
\ \ %
\endisadelimproof
\isatagproof
\isacommand{apply}\isamarkupfalse%
\ {\isacharparenleft}auto\ simp\ add{\isacharcolon}list{\isacharunderscore}copy{\isacharunderscore}def{\isacharparenright}\isanewline
\ \ \isacommand{apply}\isamarkupfalse%
\ {\isacharparenleft}subgoal{\isacharunderscore}tac\ {\isachardoublequoteopen}min\ {\isacharparenleft}length\ xs{\isacharparenright}\ n\ {\isacharplus}\ min\ {\isacharparenleft}length\ ys\ {\isacharminus}\ m{\isacharparenright}\ l\ {\isacharequal}\ n{\isacharplus}l{\isachardoublequoteclose}{\isacharparenright}\isanewline
\ \ \ \isacommand{by}\isamarkupfalse%
\ auto
\endisatagproof
\end{isabellebody}
Procedure abstractions at Simpl level to Isabelle/HOL level is very useful when a piece of code can be assumed correct because in this way we just need to deal with several simple lemmas. Otherwise, the verification condition generator will create many complex pre- and post-conditions to prove.

\subsubsection{Deep and shallow copy}
For efficiency reasons, Java instance variables are passed as reference arguments to class methods. However, in Simpl all variables are passed by value carrying out a deep copy of the parameter, for which modifications are not returned back to the global variable when the procedure finishes. Although Simpl allows to define pointers, adding pointers would make the verification more complicated. Since Simpl allows returning of multiple variables, instead of using pointers reference-passed arguments in the Java implementation are returned variables in the specification, therefore allowing to reflect modifications over these arguments. 

%In OpenJDK's implementation, when $binary\_sort(a,\ lo,\ hi,\ start)$ is called to extend a partially sorted sub-array to the minimal length, the field variable $this.a$ is passed to the method as a parameter because to access a local variable $a$ is faster than a field variable. Both the field variable and the local variable that refer to the array are just references to an area of the heap so shallow copy happens when passing the parameter. But an assignment of type $list$ in Isabelle does deep copy every time which means no side effect will happen to the parameters passed to a procedure. Although Simpl supports pointers with a split heap model\cite{}, we solve the problem in a simpler way. After we sort a sub-array, the sorted list will be returned as a return value to the caller and we assign the return value to the original list variable. Therefore, the procedure call of binary sort is defined as:

For instance, in OpenJDK's implementation, when $binary\_sort(a,\ lo,\ hi,\ start)$ is called to extend a partially sorted sub-array to the minimal length, the field variable $this.a$ is passed to the method as a parameter because the access a local variable $a$ is faster than a field variable. In our specification, after we sort a sub-array, the sorted list will be returned to the caller and we assign the returned value to the original list variable. Therefore, the procedure call of binary sort is defined as:
\begin{isabellebody}
{\isasymacute}a\ {\isacharcolon}{\isacharequal}{\isacharequal}\ CALL\ binary{\isacharunderscore}sort{\isacharparenleft}{\isasymacute}a{\isacharcomma}\ {\isasymacute}lo{\isacharcomma}\ {\isasymacute}lo{\isacharplus}{\isasymacute}force{\isacharcomma}\ {\isasymacute}lo{\isacharplus}{\isasymacute}run{\isacharunderscore}len{\isacharunderscore}i{\isacharparenright}{\isacharsemicolon}{\isacharsemicolon}\isanewline
\end{isabellebody}

\subsubsection{Methods with bitwise operations}

There are two private methods involve bitwise operations, $minRunLength(int\ n)$ and $ensureCapacity(int minCapacity)$. 
%However, as mentioned above, we declare the variables that represent the length of arrays have type $nat$ which cannot be calculated within bitwise operations. Although there is a $word$ datatype specification in Isabelle/HOL designed to work on bitwise operations, it is complicated to transfer the data type between $nat$ and $word$ and will further entangle the process of verification. 
%We investigated the functionalities of the two methods and decide to ignore them in our Simpl procedures. 
The method $minRunLength(int n)$ will calculate a number $k$, such that $16 \leq k \leq 32$ and $n/k$ is close to, but strictly less than, an exact power of 2. This return value is the minimum acceptable run length for an array of the specified length where run is an ordered segment of the original array and will be merged later. In general, the purpose of the method is to find a suitable threshold to improve the performance. Therefore, in our Simpl implementation, we can just assign this value to the number 16. 
% and the price we pay is the performance which does not matter very much for our verification of functionality.  
Similarly, the method $ensureCapacity(int minCapacity)$ is to ensure a comparatively low extra space to be used. 
%Just as we do not care about the time consumption, the space complexity does not influence verification either. 
Therefore, we just create enough space for new arrays using $replicate$ function, which simplifies the verification. 

\subsubsection{An example of translation}

The method $mergeCollapse()$ is defined as follows in Java:
\begin{lstlisting}
private void mergeCollapse() {
        while (stackSize > 1) {
            int n = stackSize - 2;
            if (n > 0 && runLen[n-1] <= runLen[n] + runLen[n+1]) {
                if (runLen[n - 1] < runLen[n + 1])
                    n--;
                mergeAt(n);
            } else if (runLen[n] <= runLen[n + 1]) {
                mergeAt(n);
            } else {
                break; // Invariant is established
            }
        }
    }
\end{lstlisting}
We translated it to the Simpl specification as follows. The translation of assignment statements, if-branches and while-loops is straightforward. Try-catch structure is used here to create the same effect as a \textit{break} statement.
\begin{isabellebody}
\isanewline
\isacommand{procedures}\isamarkupfalse%
\ {\isacharparenleft}\isakeyword{imports}\ globals{\isacharunderscore}var{\isacharparenright} \ \ 
merge{\isacharunderscore}collapse{\isacharparenleft}{\isacharparenright}\ \ 
\isakeyword{where}\ n{\isacharcolon}{\isacharcolon}nat\ \isakeyword{in}\isanewline
{\isachardoublequoteopen}TRY\ \isanewline
\ \ WHILE\ {\isasymacute}stack{\isacharunderscore}size\ {\isachargreater}\ {\isadigit{1}}\ DO\isanewline
\ \ \ \ {\isasymacute}n\ {\isacharcolon}{\isacharequal}{\isacharequal}\ {\isasymacute}stack{\isacharunderscore}size{\isacharminus}{\isadigit{2}}{\isacharsemicolon}{\isacharsemicolon}\isanewline
\ \ \ \ IF\ {\isacharparenleft}{\isasymacute}n{\isachargreater}{\isadigit{0}}\ {\isasymand}\ {\isasymacute}run{\isacharunderscore}len{\isacharbang}{\isacharparenleft}{\isasymacute}n{\isacharminus}{\isadigit{1}}{\isacharparenright}\ {\isasymle}\ {\isasymacute}run{\isacharunderscore}len{\isacharbang}{\isasymacute}n\ {\isacharplus}\ {\isasymacute}run{\isacharunderscore}len{\isacharbang}{\isacharparenleft}{\isasymacute}n{\isacharplus}{\isadigit{1}}{\isacharparenright}{\isacharparenright}\isanewline
\ \ \ \ {\isasymor}\ {\isacharparenleft}{\isasymacute}n{\isachargreater}{\isadigit{1}}\ {\isasymand}\ {\isasymacute}run{\isacharunderscore}len{\isacharbang}{\isacharparenleft}{\isasymacute}n{\isacharminus}{\isadigit{2}}{\isacharparenright}\ {\isasymle}\ {\isasymacute}run{\isacharunderscore}len{\isacharbang}{\isacharparenleft}{\isasymacute}n{\isacharminus}{\isadigit{1}}{\isacharparenright}\ {\isacharplus}\ {\isasymacute}run{\isacharunderscore}len{\isacharbang}{\isasymacute}n{\isacharparenright}\ THEN\ \isanewline
\ \ \ \ \ \ IF\ {\isasymacute}run{\isacharunderscore}len{\isacharbang}{\isacharparenleft}{\isasymacute}n{\isacharminus}{\isadigit{1}}{\isacharparenright}\ {\isacharless}\ {\isasymacute}run{\isacharunderscore}len{\isacharbang}{\isacharparenleft}{\isasymacute}n{\isacharplus}{\isadigit{1}}{\isacharparenright}\ THEN\ \isanewline
\ \ \ \ \ \ \ \ {\isasymacute}n{\isacharcolon}{\isacharequal}{\isacharequal}{\isasymacute}n{\isacharminus}{\isadigit{1}}\ \isanewline
\ \ \ \ \ \ FI\isanewline
\ \ \ \ ELSE\ \isanewline
\ \ \ \ \ \ IF\ {\isasymacute}n{\isacharless}{\isadigit{0}}\ {\isasymor}\ {\isasymacute}run{\isacharunderscore}len{\isacharbang}{\isasymacute}n\ {\isachargreater}\ {\isasymacute}run{\isacharunderscore}len{\isacharbang}{\isacharparenleft}{\isasymacute}n{\isacharplus}{\isadigit{1}}{\isacharparenright}\ THEN\ \isanewline
\ \ \ \ \ \ \ \ THROW\ \isanewline
\ \ \ \ \ \ FI\isanewline
\ \ \ \ FI{\isacharsemicolon}{\isacharsemicolon}\isanewline
\ \ \ \ CALL\ merge{\isacharunderscore}at{\isacharparenleft}{\isasymacute}n{\isacharparenright}\ \isanewline
\ \ OD\ \isanewline
CATCH\ SKIP\ END{\isachardoublequoteclose}\isanewline
\end{isabellebody}

% * <zhaoyongwang@gmail.com> 2018-07-02T01:24:13.921Z:
% 
% 这部分，要给一个Simpl代码的整体结构，扩展突出一些Timsort的Simpl代码里面比较复杂的例子，现在只有merge-collapse这个例子。
% 
% ^.
\section{Formal Verification by Hoare Logic}

For formal verification, a large number of aspects can be verified, such as functional correctness, sorting stability and absence of illegal array indexes. In our work, we mainly focus on the stack invariant which is the most important feature in Timsort. Gouw et al.\cite{deGouw2015OpenJDKsCase} have found a bug that breaks the invariant in OpenJDK's implementation. In this section, we start from the broken invariant of the stack and then prove the invariants of stack relevant and irrelevant procedures.

% * <zhaoyongwang@gmail.com> 2018-07-02T01:39:25.881Z:
% 
% "In our work, we mainly focus on the absence of the bug reported by Stijn de Gouw"，这句把我们这篇文章的价值降低不少。不从这个角度来说。 我们目前主要验证不变式，和部分代码的functional correctness，下一步验证完全的functional correctness。
% 
% ^ <yu.zhang.7161@student.lu.se> 2018-07-08T15:18:19.921Z.
\subsection{Broken Invariant}

In Timsort, a collection of sorted pieces of the array are maintained in a stack. These pieces are kept in the pattern that the length of each piece is larger than the length of its next piece and is the sum of the lengths of its next two pieces if it has. These two rules ensure the efficiency of merge sort. Moreover, as we mentioned above, another rule is that the minimal length of sorted piece is 16 in our implementation. Timsort implentation refers these sorted piece as \textit{runs}, so we will call them as runs in the following part of this section. Besides, we will call these rules about the stack as stack invariant. With the stack invariant, it is possible to infer that the space needed by the stack is a fixed number and can be calculated immediately given the length of the array to sort. Based on this property, the space of stack can be allocated in initialization as OpenJDK does. The following piece of code show the constructed function in Timsort class.
\begin{lstlisting}
int stackLen = (len < 120 ? 5 :
                len < 1542 ? 10 :
                len < 119151 ? 19 : 40);
    runBase = new int[stackLen];
    runLen = new int[stackLen];
\end{lstlisting}
If the length of array is less than 120, then 5 elements is enough for the stack. Actually, 4 is already a safe bound and the bounds in OpenJDK are slightly adjusted for more safety. Similarly, if the length of array is less than 1542, then 9 is the safe boundary. The default value is 40, because of the maximal value of integer is $2^{31}-1$ in Java. 
%We can use Isabelle to verify these conclusion. We first show the following lemma:
%\begin{isabellebody}
%\isacommand{lemma}\isamarkupfalse%
%\ {\isachardoublequoteopen}elem{\isacharunderscore}inv\ rl\ {\isadigit{0}}\ u\ {\isasymLongrightarrow}
%\ elem{\isacharunderscore}inv\ rl\ {\isadigit{1}}\ u\ {\isasymLongrightarrow}\ \isanewline
%elem{\isacharunderscore}bigger{\isacharunderscore}than{\isacharunderscore}next\ rl\ {\isadigit{2}}\ {\isasymLongrightarrow}\ 
%elem{\isacharunderscore}larger{\isacharunderscore}than{\isacharunderscore}bound\ rl\ {\isadigit{3}}\ u\ {\isasymLongrightarrow}\isanewline
%length\ rl\ {\isacharequal}\ {\isadigit{4}}\ {\isasymLongrightarrow}\ u\ {\isasymge}\ {\isadigit{1}}{\isadigit{6}}\ 
%{\isasymLongrightarrow}\ sum\ rl\ {\isasymge}\ {\isadigit{1}}{\isadigit{1}}{\isadigit{9}}{\isachardoublequoteclose}\isanewline
%
%\end{isabellebody}

Then, for fixed number of stack length $l$, we can verify that it is safe if the array length is no longer than $f(l)=u*(fib(l+1)-1)+((fib2(l+1))-(l+1))$ where $fib$ and $fib2$ are fibonacci series and modified fibonacci series.
\begin{isabellebody}
\isacommand{fun}\isamarkupfalse%
\ fib{\isacharcolon}{\isacharcolon}\ {\isachardoublequoteopen}nat\ {\isasymRightarrow}\ nat{\isachardoublequoteclose}\ \isakeyword{where}\isanewline
{\isachardoublequoteopen}fib\ {\isadigit{0}}\ {\isacharequal}\ {\isadigit{1}}{\isachardoublequoteclose}\ {\isacharbar}\isanewline
{\isachardoublequoteopen}fib\ {\isacharparenleft}Suc\ {\isadigit{0}}{\isacharparenright}\ {\isacharequal}\ {\isadigit{1}}{\isachardoublequoteclose}\ {\isacharbar}\isanewline
{\isachardoublequoteopen}fib\ {\isacharparenleft}Suc\ {\isacharparenleft}Suc\ n{\isacharparenright}{\isacharparenright}\ {\isacharequal}\ fib{\isacharparenleft}n{\isacharparenright}\ {\isacharplus}\ fib{\isacharparenleft}Suc\ n{\isacharparenright}{\isachardoublequoteclose}\isanewline
\isanewline
\isacommand{fun}\isamarkupfalse%
\ fib{\isadigit{2}}{\isacharcolon}{\isacharcolon}\ {\isachardoublequoteopen}nat\ {\isasymRightarrow}\ nat{\isachardoublequoteclose}\ \isakeyword{where}\isanewline
{\isachardoublequoteopen}fib{\isadigit{2}}\ {\isadigit{0}}\ {\isacharequal}\ {\isadigit{0}}{\isachardoublequoteclose}\ {\isacharbar}\isanewline
{\isachardoublequoteopen}fib{\isadigit{2}}\ {\isacharparenleft}Suc\ {\isadigit{0}}{\isacharparenright}\ {\isacharequal}\ {\isadigit{1}}{\isachardoublequoteclose}\ {\isacharbar}\isanewline
{\isachardoublequoteopen}fib{\isadigit{2}}\ {\isacharparenleft}Suc\ {\isacharparenleft}Suc\ n{\isacharparenright}{\isacharparenright}\ {\isacharequal}\ fib{\isadigit{2}}{\isacharparenleft}n{\isacharparenright}\ {\isacharplus}\ fib{\isadigit{2}}{\isacharparenleft}Suc\ n{\isacharparenright}\ {\isacharplus}\ {\isadigit{1}}{\isachardoublequoteclose}\isanewline
\isanewline
\end{isabellebody}
We prove $f(l)$ is the safe bound by showing that in the worst case the exact length of array is $f(l)$. The worst case means that each element in the $run\_len$ stack is exactly the least value it can be. So in the worst case, the sum of a full $run\_len$ stack is the least length of the array to be sorted by definition of $run\_len$. Consequently, arrays with less elements than the least length are safe because they will not cause the stack overflow. So we prove that the least value of each element is $u*fib(k) + fib2(k)$ where k is the index of the elements.
\begin{isabellebody}
\isacommand{lemma}\isamarkupfalse%
\ run{\isacharunderscore}len{\isacharunderscore}elem{\isacharunderscore}lower{\isacharunderscore}bound{\isacharcolon}\isanewline {\isachardoublequoteopen}
{\isasymforall}i{\isachardot}\ {\isadigit{3}}{\isasymle}i\ {\isasymand}\ i{\isasymle}l\ {\isasymlongrightarrow}\ elem{\isacharunderscore}inv\ rl\ {\isacharparenleft}l{\isacharminus}i{\isacharparenright}\ u\ {\isasymLongrightarrow}\isanewline
elem{\isacharunderscore}bigger{\isacharunderscore}than{\isacharunderscore}next\ rl\ {\isacharparenleft}l{\isacharminus}{\isadigit{2}}{\isacharparenright}\ {\isasymLongrightarrow}\isanewline
elem{\isacharunderscore}larger{\isacharunderscore}than{\isacharunderscore}bound\ rl\ {\isacharparenleft}l{\isacharminus}{\isadigit{1}}{\isacharparenright}\ u\ {\isasymLongrightarrow}\ length\ rl\ {\isacharequal}\ l\ {\isasymLongrightarrow}\ l{\isasymge}{\isadigit{2}}\ {\isasymLongrightarrow}\ k{\isacharless}l\ \isanewline
{\isasymLongrightarrow}\ rl{\isacharbang}{\isacharparenleft}l{\isacharminus}{\isadigit{1}}{\isacharminus}k{\isacharparenright}\ {\isasymge}\ u{\isacharasterisk}{\isacharparenleft}fib\ k{\isacharparenright}\ {\isacharplus}\ {\isacharparenleft}fib{\isadigit{2}}\ k{\isacharparenright}{\isachardoublequoteclose}\isanewline
\end{isabellebody}
As a result of the lower bound of every elements, we can conclude that the sum of $run\_len$ stack is $f(l)$.
\begin{isabellebody}
\isacommand{lemma}\isamarkupfalse%
\ run{\isacharunderscore}len{\isacharunderscore}sum{\isacharunderscore}lower{\isacharunderscore}bound{\isacharcolon}\ {\isachardoublequoteopen}\isanewline
{\isasymforall}i{\isachardot}\ {\isadigit{3}}{\isasymle}i\ {\isasymand}\ i{\isasymle}l\ {\isasymlongrightarrow}\ elem{\isacharunderscore}inv\ rl\ {\isacharparenleft}l{\isacharminus}i{\isacharparenright}\ u\ {\isasymLongrightarrow}\isanewline
elem{\isacharunderscore}bigger{\isacharunderscore}than{\isacharunderscore}next\ rl\ {\isacharparenleft}l{\isacharminus}{\isadigit{2}}{\isacharparenright}\ {\isasymLongrightarrow}\ \isanewline
elem{\isacharunderscore}larger{\isacharunderscore}than{\isacharunderscore}bound\ rl\ {\isacharparenleft}l{\isacharminus}{\isadigit{1}}{\isacharparenright}\ u\ {\isasymLongrightarrow}\ length\ rl\ {\isacharequal}\ l\ {\isasymLongrightarrow}\ l{\isasymge}{\isadigit{2}}\isanewline
{\isasymLongrightarrow}\ sumn\ rl\ l\ {\isasymge}\ u{\isacharasterisk}{\isacharparenleft}{\isacharparenleft}fib\ {\isacharparenleft}l{\isacharplus}{\isadigit{1}}{\isacharparenright}{\isacharparenright}{\isacharminus}{\isadigit{1}}{\isacharparenright}\ {\isacharplus}\ {\isacharparenleft}{\isacharparenleft}fib{\isadigit{2}}\ {\isacharparenleft}l{\isacharplus}{\isadigit{1}}{\isacharparenright}{\isacharparenright}\ {\isacharminus}\ {\isacharparenleft}l{\isacharplus}{\isadigit{1}}{\isacharparenright}{\isacharparenright}{\isachardoublequoteclose}\isanewline
\end{isabellebody}
In order to maintain the stack invariant, every time after a new run is pushed onto the top of stack, the method $mergeCollapse()$ will be called to check whether the stack invariant holds. If it does not, then two continuous runs will be merged to a larger run and then loop over until the stack invariant holds. However, the termination condition in the old version $mergeCollaps()$ is not strong enough to make sure that the stack invariants hold for all the elements in the stack. The result is that we cannot conclude that the space allocated to the stack in the constructor function is adequate and a runtime error might happen. As a counter example, which breaks the stack invariant, and the worst case, which does cause the index out of bound exception, is given in \cite{deGouw2015OpenJDKsCase} together with the fixed version of $mergeCollapse()$. We express the new version along with other relative methods using Simpl in Isabelle/HOL and verify that the stack invariant does hold after $newMergeCollapse$ ensuring that an implementation of our specification is correct and thus does not trigger the error.

\subsection{Prove Stack Invariant Relevant Procedures}

To prove stack invariants, we could only consider the procedures that modify the elements in the stack. These procedures in Timsort are $pushRun(int\ runBase, int\ runLen)$, $mergeCollapse()$, $mergeForceCollapse()$ and $mergeAt(int\ i)$. This subsection discusses the invariant proof of them. 
%In the following picture, we show the call graph in the Timsort implementation. What we focus on is the properties associated with the stack invariant. Therefore, we only need to take special care to the procedures that modify the elements in stack, namely $pushRun(int\ runBase, int\ runLen)$, $mergeCollapse()$, $mergeForceCollapse()$ and $mergeAt(int\ i)$.\\

Because most of work on the stack invariant has been done in KeY\cite{deGouw2015OpenJDKsCase}, we adopt them into Isabelle and prove them on Timsort procedures specified in Simpl. There are several properties that must be satisfied through the execution of the program. For example, given the length of stack is $l$, the index points to the top of stack should satisfy the property $index \ge 0 \wedge index \le l$. In KeY, these properties that hold all the time are called class invariant and are annotated by the keyword "invariant" so that they do not need to be added manually to the pre- and post-conditions of each procedure. In Simpl, we could define the pre- and post-conditions that imply the invariant. The invariant is defined in Isabelle as follows.

%However, in Isabelle there is not the concept of class and we have explicitly add these invariant in every specification. Nevertheless, this is not too burdensome because we can use definition to abbreviate the invariant. The following code is the definition of invariant and we just need to add the name of definition in specification is enough.
\begin{isabellebody}
\isanewline\isacommand{definition}\isamarkupfalse%
\ invariant\ {\isacharcolon}{\isacharcolon}\ {\isachardoublequoteopen}nat\ list\ {\isasymRightarrow}\ nat\ list\ {\isasymRightarrow}\ int\ list\ {\isasymRightarrow}\ nat\ {\isasymRightarrow}\ bool{\isachardoublequoteclose}\isanewline \isakeyword{where}\ 
{\isachardoublequoteopen}invariant\ run{\isacharunderscore}len\ run{\isacharunderscore}base\ a\ stack{\isacharunderscore}size\ {\isasymequiv}\ \isanewline
{\isacharparenleft}size\ {\isacharparenleft}run{\isacharunderscore}base{\isacharparenright}\ {\isacharequal}\ size\ {\isacharparenleft}run{\isacharunderscore}len{\isacharparenright}{\isacharparenright}\ {\isasymand}\isanewline
{\isacharparenleft}size\ a\ {\isacharless}\ {\isadigit{1}}{\isadigit{2}}{\isadigit{0}}\ {\isasymlongrightarrow}\ \ size\ run{\isacharunderscore}len\ {\isacharequal}\ {\isadigit{4}}{\isacharparenright}\ {\isasymand}\isanewline
{\isacharparenleft}size\ a\ {\isasymge}\ {\isadigit{1}}{\isadigit{2}}{\isadigit{0}}\ {\isasymand}\ size\ a\ {\isacharless}\ {\isadigit{1}}{\isadigit{5}}{\isadigit{4}}{\isadigit{2}}\ {\isasymlongrightarrow}\ \ size\ run{\isacharunderscore}len\ {\isacharequal}\ {\isadigit{9}}{\isacharparenright}\ {\isasymand}\isanewline
{\isacharparenleft}size\ a\ {\isasymge}{\isadigit{1}}{\isadigit{5}}{\isadigit{4}}{\isadigit{2}}\ {\isasymand}\ size\ a\ {\isacharless}\ {\isadigit{1}}{\isadigit{1}}{\isadigit{9}}{\isadigit{1}}{\isadigit{5}}{\isadigit{1}}\ {\isasymlongrightarrow}\ \ size\ run{\isacharunderscore}len\ {\isacharequal}\ {\isadigit{1}}{\isadigit{8}}{\isacharparenright}\ {\isasymand}\isanewline
{\isacharparenleft}size\ a\ {\isasymge}\ {\isadigit{1}}{\isadigit{1}}{\isadigit{9}}{\isadigit{1}}{\isadigit{5}}{\isadigit{1}}\ {\isasymand}\ size\ a\ {\isacharless}\ {\isadigit{2}}{\isadigit{9}}{\isadigit{1}}{\isadigit{7}}{\isadigit{1}}{\isadigit{9}}{\isadigit{6}}{\isadigit{4}}{\isadigit{9}}{\isadigit{6}}\ {\isasymlongrightarrow}\ \ size\ run{\isacharunderscore}len\ {\isacharequal}\ {\isadigit{3}}{\isadigit{9}}{\isacharparenright}\ {\isasymand}\isanewline
{\isacharparenleft}size\ a\ {\isacharless}\ {\isadigit{2}}{\isadigit{9}}{\isadigit{1}}{\isadigit{7}}{\isadigit{1}}{\isadigit{9}}{\isadigit{6}}{\isadigit{4}}{\isadigit{9}}{\isadigit{6}}{\isacharparenright}\ {\isasymand}\isanewline
{\isacharparenleft}run{\isacharunderscore}base{\isacharbang}{\isadigit{0}}\ {\isacharplus}\ {\isacharparenleft}sumn\ run{\isacharunderscore}len\ stack{\isacharunderscore}size{\isacharparenright}\ {\isasymle}\ size\ a{\isacharparenright}\ {\isasymand}\isanewline
{\isacharparenleft}stack{\isacharunderscore}size\ {\isasymge}\ {\isadigit{0}}{\isacharparenright}\ {\isasymand}\ {\isacharparenleft}stack{\isacharunderscore}size\ {\isasymle}\ size\ run{\isacharunderscore}base{\isacharparenright}\ {\isasymand}\isanewline
{\isacharparenleft}{\isasymforall}i{\isachardot}\ {\isacharparenleft}{\isacharparenleft}i{\isasymge}{\isadigit{5}}\ {\isasymand}\ i{\isasymle}stack{\isacharunderscore}size{\isacharparenright}{\isasymlongrightarrow}{\isacharparenleft}elem{\isacharunderscore}inv\ run{\isacharunderscore}len\ {\isacharparenleft}stack{\isacharunderscore}size{\isacharminus}i{\isacharparenright}\ {\isadigit{1}}{\isadigit{6}}{\isacharparenright}{\isacharparenright}{\isacharparenright}\ {\isasymand}\isanewline
{\isacharparenleft}stack{\isacharunderscore}size\ {\isasymge}{\isadigit{4}}\ {\isasymlongrightarrow}\ elem{\isacharunderscore}bigger{\isacharunderscore}than{\isacharunderscore}next\ run{\isacharunderscore}len\ {\isacharparenleft}stack{\isacharunderscore}size{\isacharminus}{\isadigit{4}}{\isacharparenright}{\isacharparenright}\ {\isasymand}\isanewline
{\isacharparenleft}stack{\isacharunderscore}size\ {\isasymge}{\isadigit{3}}\ {\isasymlongrightarrow}\ elem{\isacharunderscore}larger{\isacharunderscore}than{\isacharunderscore}bound\ run{\isacharunderscore}len\ {\isacharparenleft}stack{\isacharunderscore}size{\isacharminus}{\isadigit{3}}{\isacharparenright}\ {\isadigit{1}}{\isadigit{6}}{\isacharparenright}\ {\isasymand}\isanewline
{\isacharparenleft}stack{\isacharunderscore}size\ {\isasymge}{\isadigit{2}}\ {\isasymlongrightarrow}elem{\isacharunderscore}larger{\isacharunderscore}than{\isacharunderscore}bound\ run{\isacharunderscore}len\ {\isacharparenleft}stack{\isacharunderscore}size{\isacharminus}{\isadigit{2}}{\isacharparenright}\ {\isadigit{1}}{\isadigit{6}}{\isacharparenright}\ {\isasymand}\isanewline
{\isacharparenleft}stack{\isacharunderscore}size\ {\isasymge}{\isadigit{1}}\ {\isasymlongrightarrow}elem{\isacharunderscore}larger{\isacharunderscore}than{\isacharunderscore}bound\ run{\isacharunderscore}len\ {\isacharparenleft}stack{\isacharunderscore}size{\isacharminus}{\isadigit{1}}{\isacharparenright}\ {\isadigit{1}}{\isacharparenright}\ {\isasymand}\ \isanewline
{\isacharparenleft}{\isasymforall}i{\isachardot}\ {\isacharparenleft}{\isacharparenleft}i{\isasymge}{\isadigit{0}}\ {\isasymand}\ i{\isacharless}stack{\isacharunderscore}size{\isacharminus}{\isadigit{1}}{\isacharparenright}{\isasymlongrightarrow}{\isacharparenleft}run{\isacharunderscore}base{\isacharbang}i\ {\isacharplus}\ run{\isacharunderscore}len{\isacharbang}i\ {\isacharequal}\ run{\isacharunderscore}base{\isacharbang}{\isacharparenleft}i{\isacharplus}{\isadigit{1}}{\isacharparenright}{\isacharparenright}{\isacharparenright}{\isacharparenright}\isanewline {\isasymand}\ 
{\isacharparenleft}elem{\isacharunderscore}larger{\isacharunderscore}than{\isacharunderscore}bound\ run{\isacharunderscore}base\ {\isadigit{0}}\ {\isadigit{0}}{\isacharparenright}{\isachardoublequoteclose}\isanewline
\end{isabellebody}

%These invariants are more thorough than the stack invariant we mentioned before.  
The invariant actually covers all aspects of the two stacks that store the indexes and lengths of $run$s. The sizes of the two stacks must be the same. The size is an exact integer number given the length of input array which is not longer than $2^{31}$. The sum of all the lengths in $stack\_len$ plus the index of first $run$ cannot exceed the length of input array. $stack\_size$ always points to the top of stack and can never be greater than the size of stack. After a merge or a new element pushed into stack, the stack invariant may be broken temporarily, so these invariants are actually looser than the stack invariant. Finally, the property that the index of each $run$ plus its length should be the index of next $run$ for every valid run stored in the stacks. 

Next, we discuss the functional correctness of Timsort procedures, where the invariant defined above is included in the pre- and post-condition of each procedure. 

The functional correctness of the procedure $push\_run$ is defined as pre- and post-conditions as follows. In Simpl, the variables with $\sigma$ on up left means the old variable in the state in the pre-condition. 
The procedure $push\_run$ is called every time when the information of a new $run$ is about to be stored in stacks. Before it is called the stack invariant should be satisfied strictly. Moreover, for a valid $run$ it is supposed to have a positive index and length and its index should be its predecessor's index plus the length if it has predecessor. Also, this $run$ cannot exceed the input array. These constrains compose the precondition of $push\_run$. After the execution of $push\_run$, not only the basic functionalities should be met, we also expect that the other elements are not changed in the procedure. 
\begin{isabellebody}
\isacommand{lemma}\isamarkupfalse%
\ {\isacharparenleft}\isakeyword{in}\ push{\isacharunderscore}run{\isacharunderscore}impl{\isacharparenright}\ push{\isacharunderscore}run{\isacharunderscore}spec{\isacharcolon}\isanewline
{\isachardoublequoteopen}{\isasymforall}{\isasymsigma}{\isachardot}\ {\isasymGamma}{\isasymturnstile}\ {\isasymlbrace}{\isasymsigma}{\isachardot}\
{\isacharparenleft}{\isasymacute}run{\isacharunderscore}len{\isacharunderscore}i{\isachargreater}{\isadigit{0}}{\isacharparenright}{\isasymand}{\isacharparenleft}{\isasymacute}run{\isacharunderscore}len{\isacharunderscore}i{\isasymle}size\ {\isasymacute}a{\isacharparenright}{\isasymand}{\isacharparenleft}{\isasymacute}run{\isacharunderscore}base{\isacharunderscore}i{\isasymge}{\isadigit{0}}{\isacharparenright}{\isasymand}\isanewline
{\isacharparenleft}{\isasymacute}stack{\isacharunderscore}size{\isachargreater}{\isadigit{0}}\isanewline
\ \ \ \ {\isasymlongrightarrow}\ {\isacharparenleft}{\isasymacute}run{\isacharunderscore}base{\isacharunderscore}i{\isacharequal}{\isasymacute}run{\isacharunderscore}base{\isacharbang}{\isacharparenleft}{\isasymacute}stack{\isacharunderscore}size{\isacharminus}{\isadigit{1}}{\isacharparenright}{\isacharplus}{\isasymacute}run{\isacharunderscore}len{\isacharbang}{\isacharparenleft}{\isasymacute}stack{\isacharunderscore}size{\isacharminus}{\isadigit{1}}{\isacharparenright}{\isacharparenright}{\isacharparenright}\isanewline
 {\isasymand} {\isacharparenleft}{\isasymacute}run{\isacharunderscore}len{\isacharunderscore}i{\isacharplus}{\isasymacute}run{\isacharunderscore}base{\isacharunderscore}i\ {\isasymle}\ size\ {\isasymacute}a{\isacharparenright}\ {\isasymand}\isanewline
{\isacharparenleft}{\isasymforall}i{\isachardot}\ {\isacharparenleft}{\isacharparenleft}i{\isasymge}{\isadigit{3}}\ {\isasymand}\ i{\isasymle}{\isasymacute}stack{\isacharunderscore}size{\isacharparenright}{\isasymlongrightarrow}{\isacharparenleft}elem{\isacharunderscore}inv\ {\isasymacute}run{\isacharunderscore}len\ {\isacharparenleft}{\isasymacute}stack{\isacharunderscore}size{\isacharminus}i{\isacharparenright}\ {\isadigit{1}}{\isadigit{6}}{\isacharparenright}{\isacharparenright}{\isacharparenright}\ {\isasymand}\isanewline
{\isacharparenleft}{\isasymacute}stack{\isacharunderscore}size{\isasymge}{\isadigit{2}}\ {\isasymlongrightarrow}\ elem{\isacharunderscore}bigger{\isacharunderscore}than{\isacharunderscore}next\ {\isasymacute}run{\isacharunderscore}len\ {\isacharparenleft}{\isasymacute}stack{\isacharunderscore}size{\isacharminus}{\isadigit{2}}{\isacharparenright}{\isacharparenright}\ {\isasymand}\isanewline
{\isacharparenleft}{\isasymacute}stack{\isacharunderscore}size{\isasymge}{\isadigit{1}}\ {\isasymlongrightarrow}\ elem{\isacharunderscore}larger{\isacharunderscore}than{\isacharunderscore}bound\ {\isasymacute}run{\isacharunderscore}len\ {\isacharparenleft}{\isasymacute}stack{\isacharunderscore}size{\isacharminus}{\isadigit{1}}{\isacharparenright}\ {\isadigit{1}}{\isadigit{6}}{\isacharparenright}\ {\isasymand}\isanewline
{\isacharparenleft}{\isasymacute}stack{\isacharunderscore}size{\isasymge}{\isadigit{0}}\ {\isasymand}\ {\isasymacute}stack{\isacharunderscore}size{\isasymle}{\isacharparenleft}size\ {\isasymacute}run{\isacharunderscore}len{\isacharparenright}{\isacharparenright}\ {\isasymand}\isanewline
{\isacharparenleft}invariant\ {\isasymacute}run{\isacharunderscore}len\ {\isasymacute}run{\isacharunderscore}base\ {\isasymacute}a\ {\isasymacute}stack{\isacharunderscore}size{\isacharparenright}\ {\isasymrbrace}\isanewline
PROC\ push{\isacharunderscore}run{\isacharparenleft}{\isasymacute}run{\isacharunderscore}base{\isacharunderscore}i{\isacharcomma}\ {\isasymacute}run{\isacharunderscore}len{\isacharunderscore}i{\isacharparenright}\ \isanewline
{\isasymlbrace}\ {\isacharparenleft}{\isasymacute}run{\isacharunderscore}base{\isacharbang}{\isacharparenleft}\isactrlbsup {\isasymsigma}\isactrlesup stack{\isacharunderscore}size{\isacharparenright}\ {\isacharequal}\ {\isasymacute}run{\isacharunderscore}base{\isacharunderscore}i{\isacharparenright}\ {\isasymand}\isanewline
{\isacharparenleft}{\isasymacute}run{\isacharunderscore}len{\isacharbang}{\isacharparenleft}\ \isactrlbsup {\isasymsigma}\isactrlesup stack{\isacharunderscore}size{\isacharparenright}\ {\isacharequal}\ {\isasymacute}run{\isacharunderscore}len{\isacharunderscore}i{\isacharparenright}\ {\isasymand}
{\isacharparenleft}{\isasymacute}stack{\isacharunderscore}size\ {\isacharequal}\ \ \isactrlbsup {\isasymsigma}\isactrlesup stack{\isacharunderscore}size\ {\isacharplus}\ {\isadigit{1}}{\isacharparenright}\isanewline
{\isasymand} {\isacharparenleft}{\isasymacute}run{\isacharunderscore}base{\isacharunderscore}i\ {\isacharequal}\ \isactrlbsup {\isasymsigma}\isactrlesup run{\isacharunderscore}base{\isacharunderscore}i{\isacharparenright}\ {\isasymand}
{\isacharparenleft}{\isasymacute}run{\isacharunderscore}len{\isacharunderscore}i\ {\isacharequal}\ \isactrlbsup {\isasymsigma}\isactrlesup run{\isacharunderscore}len{\isacharunderscore}i{\isacharparenright}\ {\isasymand}\isanewline
{\isacharparenleft}{\isasymforall}i{\isachardot}\ {\isacharparenleft}i{\isasymge}{\isadigit{0}}\ {\isasymand}\ i{\isacharless}{\isasymacute}stack{\isacharunderscore}size{\isacharminus}{\isadigit{1}}{\isacharparenright}\ {\isasymlongrightarrow}\ {\isacharparenleft}{\isasymacute}run{\isacharunderscore}len{\isacharbang}i\ {\isacharequal}\ \isactrlbsup {\isasymsigma}\isactrlesup run{\isacharunderscore}len{\isacharbang}i{\isacharparenright}{\isacharparenright}\ {\isasymand}\isanewline
{\isacharparenleft}{\isasymforall}i{\isachardot}\ {\isacharparenleft}i{\isasymge}{\isadigit{0}}\ {\isasymand}\ i{\isacharless}{\isasymacute}stack{\isacharunderscore}size{\isacharminus}{\isadigit{1}}{\isacharparenright}\ {\isasymlongrightarrow}\ {\isacharparenleft}{\isasymacute}run{\isacharunderscore}base{\isacharbang}i\ {\isacharequal}\ \isactrlbsup {\isasymsigma}\isactrlesup run{\isacharunderscore}base{\isacharbang}i{\isacharparenright}{\isacharparenright}\ {\isasymand}\ \isanewline
{\isacharparenleft}invariant\ {\isasymacute}run{\isacharunderscore}len\ {\isasymacute}run{\isacharunderscore}base\ {\isasymacute}a\ {\isasymacute}stack{\isacharunderscore}size{\isacharparenright}\ {\isasymand}
{\isacharparenleft}size\ {\isasymacute}a\ {\isacharequal}\ size\ \isactrlbsup {\isasymsigma}\isactrlesup a{\isacharparenright}\ {\isasymrbrace}{\isachardoublequoteclose}
\end{isabellebody}

Now, we discuss the procedure $merge\_at$. The pre- and post-conditions of the procedure is defined as follows. 
For the top three elements in $stack\_len$, $X$,$Y$ and $Z$, there are different merge strategies for different cases. It can either happen on $X$ and $Y$ or on $Y$ and $Z$ given different conditions. So the index may either point at $X$ or at $Y$ if there are at least three elements. After the merge, except for basic functionalities, we can also prove that the sum of valid elements in $stack\_len$ does not change, which means merge does not change the total number of elements. Moreover, after merge we expect that the last element in $stack\_len$ does not decrease, which mean the process of merge can either make the last valid $run$ longer or stay it unchanged.
\begin{isabellebody}
\isacommand{lemma}\isamarkupfalse%
\ {\isacharparenleft}\isakeyword{in}\ merge{\isacharunderscore}at{\isacharunderscore}impl{\isacharparenright}\ merge{\isacharunderscore}at{\isacharunderscore}spec{\isacharcolon}\ \isanewline
{\isachardoublequoteopen}{\isasymforall}{\isasymsigma}{\isachardot}\ {\isasymGamma}{\isasymturnstile}\ {\isasymlbrace}{\isasymsigma}{\isachardot} 
{\isacharparenleft}{\isasymacute}stack{\isacharunderscore}size{\isacharequal}{\isadigit{2}}\ {\isasymlongrightarrow}\ {\isasymacute}i{\isacharequal}{\isasymacute}stack{\isacharunderscore}size{\isacharminus}{\isadigit{2}}{\isacharparenright}\ {\isasymand}\isanewline
{\isacharparenleft}{\isasymacute}stack{\isacharunderscore}size{\isasymge}{\isadigit{3}}\ {\isasymlongrightarrow}\ {\isacharparenleft}{\isasymacute}i{\isacharequal}{\isasymacute}stack{\isacharunderscore}size{\isacharminus}{\isadigit{2}}\ {\isasymor}{\isasymacute}i{\isacharequal}{\isasymacute}stack{\isacharunderscore}size{\isacharminus}{\isadigit{3}}{\isacharparenright}{\isacharparenright}\ {\isasymand}\isanewline
{\isacharparenleft}{\isasymacute}stack{\isacharunderscore}size\ {\isachargreater}\ {\isadigit{1}}{\isacharparenright}\ {\isasymand}\ {\isacharparenleft}{\isasymacute}i{\isasymge}{\isadigit{0}}{\isacharparenright}\ {\isasymand}
{\isacharparenleft}{\isasymacute}a\ {\isasymnoteq}\ {\isacharbrackleft}{\isacharbrackright}{\isacharparenright}\ {\isasymand}\isanewline
{\isacharparenleft}invariant\ {\isasymacute}run{\isacharunderscore}len\ {\isasymacute}run{\isacharunderscore}base\ {\isasymacute}a\ {\isasymacute}stack{\isacharunderscore}size{\isacharparenright}\ {\isasymrbrace}\ \isanewline
\ \ PROC\ merge{\isacharunderscore}at{\isacharparenleft}{\isasymacute}i{\isacharparenright}\ \isanewline
{\isasymlbrace}\ {\isacharparenleft}{\isasymacute}i\ {\isacharequal}\ \isactrlbsup {\isasymsigma}\isactrlesup i{\isacharparenright}\ {\isasymand}\ {\isacharparenleft}{\isasymacute}run{\isacharunderscore}base{\isacharbang}{\isadigit{0}}\ {\isacharequal}\ \isactrlbsup {\isasymsigma}\isactrlesup run{\isacharunderscore}base{\isacharbang}{\isadigit{0}}{\isacharparenright}\ {\isasymand}
{\isacharparenleft}size\ {\isasymacute}a\ {\isacharequal}\ size\ \isactrlbsup {\isasymsigma}\isactrlesup a{\isacharparenright}\ {\isasymand}\isanewline
{\isacharparenleft}{\isasymacute}stack{\isacharunderscore}size\ {\isacharequal}\ \isactrlbsup {\isasymsigma}\isactrlesup stack{\isacharunderscore}size\ {\isacharminus}\ {\isadigit{1}}{\isacharparenright}\ {\isasymand}\isanewline
{\isacharparenleft}{\isasymacute}run{\isacharunderscore}len{\isacharbang}{\isasymacute}i\ {\isacharequal}\ {\isacharparenleft}\isactrlbsup {\isasymsigma}\isactrlesup run{\isacharunderscore}len{\isacharparenright}{\isacharbang}{\isacharparenleft}{\isasymacute}i{\isacharparenright}\ {\isacharplus}\ {\isacharparenleft}\isactrlbsup {\isasymsigma}\isactrlesup run{\isacharunderscore}len{\isacharbang}{\isacharparenleft}{\isasymacute}i{\isacharplus}{\isadigit{1}}{\isacharparenright}{\isacharparenright}{\isacharparenright}\ {\isasymand}\ \isanewline
{\isacharparenleft}{\isacharparenleft}{\isasymacute}i\ {\isacharequal}\ \isactrlbsup {\isasymsigma}\isactrlesup stack{\isacharunderscore}size\ {\isacharminus}{\isadigit{3}}{\isacharparenright} \isanewline
\ \ {\isasymlongrightarrow}{\isacharparenleft}{\isasymacute}run{\isacharunderscore}len{\isacharbang}{\isacharparenleft}{\isasymacute}i{\isacharplus}{\isadigit{1}}{\isacharparenright}\ {\isacharequal}\ \isactrlbsup {\isasymsigma}\isactrlesup run{\isacharunderscore}len{\isacharbang}{\isacharparenleft}{\isasymacute}i{\isacharplus}{\isadigit{2}}{\isacharparenright}{\isacharparenright}\isanewline
\ \ \ \ \ \ \  {\isasymand}\ {\isacharparenleft}{\isasymacute}run{\isacharunderscore}base{\isacharbang}{\isacharparenleft}{\isasymacute}i{\isacharplus}{\isadigit{1}}{\isacharparenright}\ {\isacharequal}\ \isactrlbsup {\isasymsigma}\isactrlesup run{\isacharunderscore}base{\isacharbang}{\isacharparenleft}{\isasymacute}i{\isacharplus}{\isadigit{2}}{\isacharparenright}{\isacharparenright}{\isacharparenright}\ {\isasymand}\isanewline
{\isacharparenleft}invariant\ {\isasymacute}run{\isacharunderscore}len\ {\isasymacute}run{\isacharunderscore}base\ {\isasymacute}a\ {\isasymacute}stack{\isacharunderscore}size{\isacharparenright}\ {\isasymand}\isanewline
{\isacharparenleft}sumn\ {\isasymacute}run{\isacharunderscore}len\ {\isasymacute}stack{\isacharunderscore}size\ {\isacharequal}\ sumn\ \isactrlbsup {\isasymsigma}\isactrlesup run{\isacharunderscore}len\ \isactrlbsup {\isasymsigma}\isactrlesup stack{\isacharunderscore}size{\isacharparenright}\ {\isasymand}\isanewline
{\isacharparenleft}{\isasymacute}run{\isacharunderscore}len{\isacharbang}{\isacharparenleft}{\isasymacute}stack{\isacharunderscore}size{\isacharminus}{\isadigit{1}}{\isacharparenright}\ {\isasymge}\ \isactrlbsup {\isasymsigma}\isactrlesup run{\isacharunderscore}len{\isacharbang}{\isacharparenleft}\isactrlbsup {\isasymsigma}\isactrlesup stack{\isacharunderscore}size{\isacharminus}{\isadigit{1}}{\isacharparenright}{\isacharparenright}\ {\isasymrbrace}{\isachardoublequoteclose}\isanewline
\end{isabellebody}

The pre- and post-conditions of the procedure $merge\_collapse$ is defined as follows.  
Every time after a new $run$ is pushed to the stack, this algorithm checks whether the stack invariant holds and merge a few sub-arrays to re-establish the stack invariant, which is what procedure $merge\_collapse$ does. Because the length of the new $run$ isnot known, the precondition only constrains the other elements in the stack invariant. But after the procedure is done, the stack invariants must strictly holds for each element. Again, there are some corollaries that come together with the postcondition just as in the specification of procedure $merge\_at$.
\begin{isabellebody}
\isacommand{lemma}\isamarkupfalse%
\ {\isacharparenleft}\isakeyword{in}\ merge{\isacharunderscore}collapse{\isacharunderscore}impl{\isacharparenright}\ merge{\isacharunderscore}collapse{\isacharunderscore}spec{\isacharcolon}\isanewline
{\isachardoublequoteopen}{\isasymforall}{\isasymsigma}{\isachardot}\ {\isasymGamma}{\isasymturnstile}\ {\isasymlbrace}{\isasymsigma}{\isachardot}\ {\isacharparenleft}{\isasymacute}stack{\isacharunderscore}size{\isachargreater}{\isadigit{0}}{\isacharparenright}\ {\isasymand}\isanewline
 {\isacharparenleft}{\isasymacute}stack{\isacharunderscore}size{\isasymge}{\isadigit{4}}\ {\isasymlongrightarrow}\ elem{\isacharunderscore}inv\ {\isasymacute}run{\isacharunderscore}len\ {\isacharparenleft}{\isasymacute}stack{\isacharunderscore}size{\isacharminus}{\isadigit{4}}{\isacharparenright}\ {\isadigit{1}}{\isadigit{6}}{\isacharparenright}\ {\isasymand}\isanewline
{\isacharparenleft}{\isasymacute}stack{\isacharunderscore}size{\isasymge}{\isadigit{3}}\ {\isasymlongrightarrow}\ elem{\isacharunderscore}bigger{\isacharunderscore}than{\isacharunderscore}next\ {\isasymacute}run{\isacharunderscore}len\ {\isacharparenleft}{\isasymacute}stack{\isacharunderscore}size{\isacharminus}{\isadigit{3}}{\isacharparenright}{\isacharparenright}\ {\isasymand}\isanewline
{\isacharparenleft}{\isasymacute}a\ {\isasymnoteq}\ {\isacharbrackleft}{\isacharbrackright}{\isacharparenright}\ {\isasymand}
{\isacharparenleft}invariant\ {\isasymacute}run{\isacharunderscore}len\ {\isasymacute}run{\isacharunderscore}base\ {\isasymacute}a\ {\isasymacute}stack{\isacharunderscore}size{\isacharparenright}\ {\isasymrbrace}\isanewline
PROC\ merge{\isacharunderscore}collapse{\isacharparenleft}{\isacharparenright}\ \isanewline
{\isasymlbrace}\ {\isacharparenleft}{\isasymforall}i{\isachardot}\ {\isacharparenleft}{\isacharparenleft}i{\isasymge}{\isadigit{3}}\ {\isasymand}\ i{\isasymle}{\isasymacute}stack{\isacharunderscore}size{\isacharparenright}{\isasymlongrightarrow}{\isacharparenleft}elem{\isacharunderscore}inv\ {\isasymacute}run{\isacharunderscore}len\ {\isacharparenleft}{\isasymacute}stack{\isacharunderscore}size{\isacharminus}i{\isacharparenright}\ {\isadigit{1}}{\isadigit{6}}{\isacharparenright}{\isacharparenright}{\isacharparenright}\isanewline
{\isasymand} {\isacharparenleft}{\isasymacute}stack{\isacharunderscore}size{\isasymge}{\isadigit{2}}\ {\isasymlongrightarrow}\ elem{\isacharunderscore}bigger{\isacharunderscore}than{\isacharunderscore}next\ {\isasymacute}run{\isacharunderscore}len\ {\isacharparenleft}{\isasymacute}stack{\isacharunderscore}size{\isacharminus}{\isadigit{2}}{\isacharparenright}{\isacharparenright}\ {\isasymand}\isanewline
{\isacharparenleft}sumn\ {\isasymacute}run{\isacharunderscore}len\ {\isasymacute}stack{\isacharunderscore}size\ {\isacharequal}\ sumn\ \isactrlbsup {\isasymsigma}\isactrlesup run{\isacharunderscore}len\ \isactrlbsup {\isasymsigma}\isactrlesup stack{\isacharunderscore}size\ {\isacharparenright}\ {\isasymand}\isanewline
{\isacharparenleft}{\isasymacute}run{\isacharunderscore}len{\isacharbang}{\isacharparenleft}{\isasymacute}stack{\isacharunderscore}size{\isacharminus}{\isadigit{1}}{\isacharparenright}\ {\isasymge}\ \isactrlbsup {\isasymsigma}\isactrlesup run{\isacharunderscore}len{\isacharbang}{\isacharparenleft}\isactrlbsup {\isasymsigma}\isactrlesup stack{\isacharunderscore}size{\isacharminus}{\isadigit{1}}{\isacharparenright}{\isacharparenright}\ {\isasymand}\isanewline
{\isacharparenleft}{\isasymacute}stack{\isacharunderscore}size{\isachargreater}{\isadigit{0}}\ {\isasymand}\ {\isasymacute}stack{\isacharunderscore}size\ {\isasymle}\ \isactrlbsup {\isasymsigma}\isactrlesup stack{\isacharunderscore}size{\isacharparenright}\ {\isasymand}\isanewline
{\isacharparenleft}invariant\ {\isasymacute}run{\isacharunderscore}len\ {\isasymacute}run{\isacharunderscore}base\ {\isasymacute}a\ {\isasymacute}stack{\isacharunderscore}size{\isacharparenright}\ {\isasymand}\isanewline
{\isacharparenleft}size\ {\isasymacute}a\ {\isacharequal}\ size\ \isactrlbsup {\isasymsigma}\isactrlesup a{\isacharparenright}{\isasymand}
{\isacharparenleft}{\isasymacute}run{\isacharunderscore}base{\isacharbang}{\isadigit{0}}\ {\isacharequal}\ \isactrlbsup {\isasymsigma}\isactrlesup run{\isacharunderscore}base{\isacharbang}{\isadigit{0}}{\isacharparenright}\ {\isasymrbrace}{\isachardoublequoteclose}\isanewline
\end{isabellebody}

Finally, for the procedure $merge\_force\_collapse$, after the entire input array has been divided into $run$s, all the $run$s are merged from back to the beginning. There is not special requirement before this procedure and we only expect that only one $run$ remains at last.
\begin{isabellebody}
\isacommand{lemma}\isamarkupfalse%
\ {\isacharparenleft}\isakeyword{in}\ merge{\isacharunderscore}force{\isacharunderscore}collapse{\isacharunderscore}impl{\isacharparenright}\ merge{\isacharunderscore}force{\isacharunderscore}collapse{\isacharunderscore}spec{\isacharcolon}\isanewline
{\isachardoublequoteopen}{\isasymforall}{\isasymsigma}{\isachardot}\ {\isasymGamma}{\isasymturnstile}\ {\isasymlbrace}{\isasymsigma}{\isachardot}\ {\isacharparenleft}{\isasymacute}stack{\isacharunderscore}size\ {\isachargreater}\ {\isadigit{0}}{\isacharparenright}\ {\isasymand}
{\isacharparenleft}{\isasymacute}a\ {\isasymnoteq}\ {\isacharbrackleft}{\isacharbrackright}{\isacharparenright}\ {\isasymand}\isanewline
{\isacharparenleft}invariant\ {\isasymacute}run{\isacharunderscore}len\ {\isasymacute}run{\isacharunderscore}base\ {\isasymacute}a\ {\isasymacute}stack{\isacharunderscore}size{\isacharparenright}\ {\isasymrbrace}\isanewline
PROC\ merge{\isacharunderscore}force{\isacharunderscore}collapse{\isacharparenleft}{\isacharparenright}\isanewline
{\isasymlbrace}\ {\isacharparenleft}{\isasymacute}stack{\isacharunderscore}size\ {\isacharequal}\ {\isadigit{1}}{\isacharparenright}\ {\isasymand}
{\isacharparenleft}invariant\ {\isasymacute}run{\isacharunderscore}len\ {\isasymacute}run{\isacharunderscore}base\ {\isasymacute}a\ {\isasymacute}stack{\isacharunderscore}size{\isacharparenright}\ {\isasymand}\isanewline
{\isacharparenleft}size\ {\isasymacute}a\ {\isacharequal}\ size\ \isactrlbsup {\isasymsigma}\isactrlesup a{\isacharparenright}\ {\isasymrbrace}{\isachardoublequoteclose}\isanewline
\end{isabellebody}

\subsection{Prove Stack Invariant Irrelevant Procedures}

For the procedures that do not modify the stack, their specification are relatively simple. We prove that their parameters are valid and the invariant that mentioned above still holds. There are several ways to prove the invariant. On one hand, we can add them in the pre- and post-conditions and prove them as normal. On the other hand, we can use the modification specification provided by Simpl instead, which describes whether a global variable is modified in this procedure. In this paper, we use the second way. For instance, the condition of the procedure $merge\_lo$ is defined as follows. 

\begin{isabellebody}
\isanewline
\isacommand{lemma}\isamarkupfalse%
\ {\isacharparenleft}\isakeyword{in}\ merge{\isacharunderscore}lo{\isacharunderscore}impl{\isacharparenright}\ merge{\isacharunderscore}lo{\isacharunderscore}modifies{\isacharcolon}\isanewline
\ \ \isakeyword{shows}\isanewline
\ \ {\isachardoublequoteopen}{\isasymforall}\ {\isasymsigma}{\isachardot}\ {\isasymGamma}{\isasymturnstile}\ {\isacharbraceleft}{\isasymsigma}{\isacharbraceright}\ PROC\ merge{\isacharunderscore}lo{\isacharparenleft}{\isasymacute}base{\isadigit{1}}{\isacharcomma}\ {\isasymacute}len{\isadigit{1}}{\isacharcomma}\ {\isasymacute}base{\isadigit{2}}{\isacharcomma}\ {\isasymacute}len{\isadigit{2}}{\isacharparenright}\isanewline
\ \ \ \ \ \ \ \ \ \ \ \ {\isacharbraceleft}t{\isachardot}\ t\ may{\isacharunderscore}only{\isacharunderscore}modify{\isacharunderscore}globals\ {\isasymsigma}\ in\ {\isacharbrackleft}a{\isacharbrackright}{\isacharbraceright}{\isachardoublequoteclose}\isanewline
\end{isabellebody}

In such a condition, the caller of $mergeLo$ knows that this procedure only modifies the array, and that the global variables that are not modified in the new state are the same as in the state before procedure call. This implicitly proves that if the related properties hold before the call, they hold after the call as well because they are not changed at all. 

\subsection{Modularity}

In the process of verification, there are several specifications involving a large number of pre- and post-conditions. As a result, it is complicated for Simpl to simplify these conditions by the verification condition generator. Moreover, if we modify only one of these conditions in the large specification, the whole proof need to be rearranged because one of the proof steps might fail after the change and the following steps will be invalid. Therefore, we need modularity to divide the large specification into several small pieces. For example, consider the specification $\Gamma,\Theta \vdash \{P\}\ c\ \{Q\}$, if both $\{P\}$and $\{ Q\}$ contain a large number of conditions, the proof size will be large consequently. However, it is possible to prove two sub-specifications  $\Gamma,\Theta \vdash \{ P\}\ c\ \{Q_1\}$ and $\Gamma,\Theta\vdash \{ P\} c \{Q_2\}$ where $\{Q_1\}\cap\{Q_2\}=\{Q\}$ and then conclude the original specification with the application of rule $HoarePartialProps.PostConjI$. 
\begin{isabellebody}
\isacommand{lemma}\isamarkupfalse%
\ PostConjI{\isacharcolon}\ \isanewline
\ \ \isakeyword{assumes}\ deriv{\isacharunderscore}Q{\isacharcolon}\ {\isachardoublequoteopen}{\isasymGamma}{\isacharcomma}{\isasymTheta}{\isasymturnstile}\isactrlbsub {\isacharslash}F\isactrlesub \ P\ c\ Q{\isacharcomma}A{\isachardoublequoteclose}\ \isanewline
\ \ \isakeyword{assumes}\ deriv{\isacharunderscore}R{\isacharcolon}\ {\isachardoublequoteopen}{\isasymGamma}{\isacharcomma}{\isasymTheta}{\isasymturnstile}\isactrlbsub {\isacharslash}F\isactrlesub \ P\ c\ R{\isacharcomma}B{\isachardoublequoteclose}\isanewline
\ \ \isakeyword{shows}\ {\isachardoublequoteopen}{\isasymGamma}{\isacharcomma}{\isasymTheta}{\isasymturnstile}\isactrlbsub {\isacharslash}F\isactrlesub \ P\ c\ {\isacharparenleft}Q\ {\isasyminter}\ R{\isacharparenright}{\isacharcomma}{\isacharparenleft}A\ {\isasyminter}\ B{\isacharparenright}{\isachardoublequoteclose}\isanewline
\end{isabellebody}
Furthermore, if only the subset $\{P_1\}$ is adequate to derive the post-conditions set $\{Q_1\}$, then we can get rid of the redundant assumptions in the first sub-specification, which can be achieved by the rule $HoarePartialDef.conseqPre$. 
\begin{isabellebody}
\isacommand{lemma}\isamarkupfalse%
\ conseqPre{\isacharcolon}\ {\isachardoublequoteopen}{\isasymGamma}{\isacharcomma}{\isasymTheta}{\isasymturnstile}\isactrlsub t\isactrlbsub {\isacharslash}F\isactrlesub \ P{\isacharprime}\ c\ Q{\isacharcomma}A\ {\isasymLongrightarrow}\ P\ {\isasymsubseteq}\ P{\isacharprime}\ {\isasymLongrightarrow}\ {\isasymGamma}{\isacharcomma}{\isasymTheta}{\isasymturnstile}\isactrlsub t\isactrlbsub {\isacharslash}F\isactrlesub \ P\ c\ Q{\isacharcomma}A{\isachardoublequoteclose}\isanewline
\end{isabellebody}
Therefore, we can reduce the size of specification to ease the verification. Another benefit is modularity, which means we can prove different aspects of algorithms in different specifications. For example, we do not need to prove correctness and stability of a sort algorithm at the same time. Instead, it is possible to verify them separately. 

\section{Evaluation and Discussion}

\subsection{Generated C code}
For this stage, we manually develop the C code according to Simpl. This process is simple because most of the Simpl statements can be directly mapped to C statements. We replace $list\_copy$ and $replicate$ with C standard library functions $memcpy$ and $malloc$ to copy an area in memory and allocate new area for arrays. Besides, we pass an array to a function body as a parameter by passing its pointer because we can only shallow copy arrays in C and therefore we avoid returning two values in function$count\_run\_and\_make\_ascending$. Further, when it comes to try-catch clauses, we translate them according to the behaviour they specify. If abrupt termination follows an assignment to return value we translate it as a return in C. Likewise, if it means to begin a new iteration or quit from iteration in a loop then it is a continue or break. In case of try-catch structure in a nested loop, we define some flag variables to tell which level of loop to break or continue avoiding use of goto statement. The other parts are straightforward. 

We tested the C implementation with simple random test cases. Randomness does not mean that each element is generated randomly, because in this case it is of slight possibility to obtain an ascending or descending array longer than the minimal length boundary for $run$s. Consequently almost all the sub-arrays are forced to extend to the minimal length. Timsort does not have an optimal performance  when all $run$s have equal length, for which basically it has no differences from the traditional merge sort with a threshold. Therefore, we generate random numbers that represent the length of ascending sub-arrays and generate random numbers increasingly in the array to obtain the test case. When it comes to check the correctness of results, it is not reliable to just traverse the result arrays and assert that every elements is less than its successor. It is also necessary to check all the elements are preserved. So the method we use is to copy the test array and call quicksort in C standard library to sort it. We assume that quicksort in C standard library is correct and conclude Timsort implementation in C is correct from the fact that two implementations generated exactly the same results. 

It is desirable to automate the translation from Simpl to C. First, for large programs it is hard to manually carry out the translation. Second, it is not reliable to generate code by hand because it is a process prone to introduce errors during the manual translation. Although there are methods to check the consistency between Simpl code and corresponding C code, it is much better if reliable C code can be generated directly. To achieve this, we expect to create a verified compiler from Simpl to C embedded in Isabelle/HOL. 

\subsection{Statistics of Specification and Proof}

We use Isabelle/HOL as the specification and verification system for our work. All derivations of our proofs have passed through the Isabelle proof kernel. We develop 400 lines of Simpl specification, which is at the same scale as those in OpenJDK. We use 1,600 lines of proof to show its correctness, most of the effort is on the stack relevant invariants. The C implementation of Timsort generated from the Simpl specification is 500 lines. %We take 2 person months for the total effort. 
Compared to a large number ($\approx$ 5000) of manual proof steps in KeY \cite{deGouw2015OpenJDKsCase}, we show a less proof effort in Simpl and the advantage of using its VCG.

\section{Conclusion and Future Work}

We have formally verified Timsort using Isabelle/HOL and Simpl. We have first specified the Timsort algorithm using the generic imperative language Simpl based on the implementation in OpenJDK. Then we have verified relevant properties w.r.t. the stack on the specification using  Simpl VCG to simplify the verification to high order logic problems that can be solved using Isabelle/HOL automated decision procedures. Finally, we have manually produced verified Timsort implementation in C from our Simpl specification.

In the future, we plan to formally verify the full functional correctness of the Timsort implementation as well as use benckmarks to fully test the performance of our Timsort implementation in C. To complete the verification framework we consider to build a verified translator from Simpl specifications to C. 
\bibliographystyle{splncs04}
\bibliography{timsort}
\end{document}